\documentclass[reprint,prl,twocolumn,showpacs,preprintnumbers,amsmath,amssymb]{revtex4}

\usepackage{graphicx}% Include figure files
\usepackage{dcolumn}% Align table columns on decimal point
\usepackage{bm}% bold math
\usepackage{multirow}
\usepackage{color}
\usepackage{amsmath}
\usepackage{mathtools}
\usepackage{color}

\DeclarePairedDelimiter\bra{\langle}{\rvert}
\DeclarePairedDelimiter\ket{\lvert}{\rangle}
\DeclarePairedDelimiterX\braket[2]{\langle}{\rangle}{#1 \delimsize\vert #2}
\DeclarePairedDelimiterX\inner[2]{\langle}{\rangle}{#1,#2}

\begin{document}
\title{Splitting between Bright and Dark excitons in Transition Metal Dichalcogenide Monolayers}

\author{J. P. Echeverry}
\author{B. Urbaszek}
\author{T. Amand}
\author{X. Marie}
\author{I. C. Gerber}
\email{igerber@insa-toulouse.fr}
\affiliation{%
Universit\'e F\'ed\'erale de Toulouse Midi Pyr\'en\'ees, INSA-CNRS-UPS, LPCNO, 135 Av. de Rangueil, 31077 Toulouse, France}

%\listfiles
%\nofiles
%\keywords{WSe2, valley dynamics, time resolved photoluminescence, transition metal dichalcogenides, two dimensional materials}
%\date{\today}

\begin{abstract}
The optical properties of transition metal dichalcogenide monolayers such as the two-dimensional semiconductors MoS$_2$ and WSe$_2$ are dominated by excitons, Coulomb bound electron-hole pairs. 
The light emission yield depends on 
whether the electron-hole transitions are optically allowed (bright) or forbidden (dark). By solving the Bethe Salpeter Equation on top of $GW$ wave functions in density functional theory calculations, we determine the sign
 and amplitude of the splitting between bright and dark exciton states. We evaluate the influence of the spin-orbit coupling on the optical spectra and clearly demonstrate the strong impact of the intra-valley Coulomb exchange 
 term on the dark-bright exciton fine structure splitting.

\end{abstract}

%\pacs{78.60.Lc,78.66.Li}

                           %Use showkeys class option if keyword
                             %display desired
                             
\maketitle

\textit{Introduction.---} Transition metal dichalcogenide (TMDC) monolayers (MLs), with chemical formula MX$_2$ , with M=Mo or W and X=S, Se or Te, are semiconductors with a direct bandgap in the visible region situated 
at the $K$-point of the Brillouin zone \cite{Mak:2010do,Splendiani:2010a}. 
In these 2D systems the broken inversion symmetry of the crystal lattice allows for optical control of the valley degree of freedom~\cite{Cao:2012a,Zeng:2012fn,Mak:2012bh,Sallen:2012a}. Combined with strong spin-orbit coupling (SOC) this leads to valley-spin locking~\cite{Xiao:2012dv}, opening exciting avenues for original applications in optoelectronics and spintronics.
Tightly bound excitons 
~\cite{Chernikov:2014ic,Ye:2014hb,He:2014bo,Wang:2015kb}, with binding energies around 0.5 eV, originate from the direct term of the electron-hole ($e$-$h)$ Coulomb interaction, enhanced by the strong 2D quantum 
confinement, the large effective masses, and the reduced dielectric screening in 2D systems. While the energy spectra of these excitons have been intensively studied by combining both experimental and theoretical technics in the last years~\cite{Qiu:2013jv,MolinaSanchez:2013hz,Ugeda:2014cy,Klots:2014gi, Qiu:2015hj, Qiu:2015fa}, little is known about their fine structure.\\
\indent For optoelectronics based on TMDC MLs it is important to clarify if the lowest energy transition is optically bright or optically dark (i.e. spin forbidden) \cite{note0}. This bright-dark exciton fine structure splitting governs the optical properties also of 2D semiconductor nano-structures such as III-V and II-VI quantum wells~\cite{Blackwood:1994vl,Puls:1997a,Amand:1997wm, Besombes:2000gk,Vanelle:2000tz} as well as CdSe nanocrystals~\cite{Crooker:2003hk}. 
In particular, the measured low photo- or electro-luminescence yield and its increase with the temperature in monolayer WSe$_2$ has been interpreted recently in terms of dark excitons lying at lower energy compared to the bright ones \cite{Arora:2015il,C5NR06782K,Wang:2015ko,Zhang:2015er,Withers:2015cf}. This will also affect the efficiency of recently demonstrated WSe$_2$ or WS$_2$ light emitting devices~\cite{Wu:2015es,Ye:2015gf}. For fundamental physics experiments an optically dark ground state can be an advantage, as it allows to study exciton quantum fluids in Bose-Einstein exciton condensates~\cite{Combescot:2007ux}.
It is thus crucial to determine both the amplitude and sign of this bright-dark exciton splitting in TMDC MLs.\\
%%%%%%%%%%%%%%%%%%%%%%%%%%%%%%%%%%%%%
\begin{figure}[htp]
\includegraphics[width=0.49\textwidth]{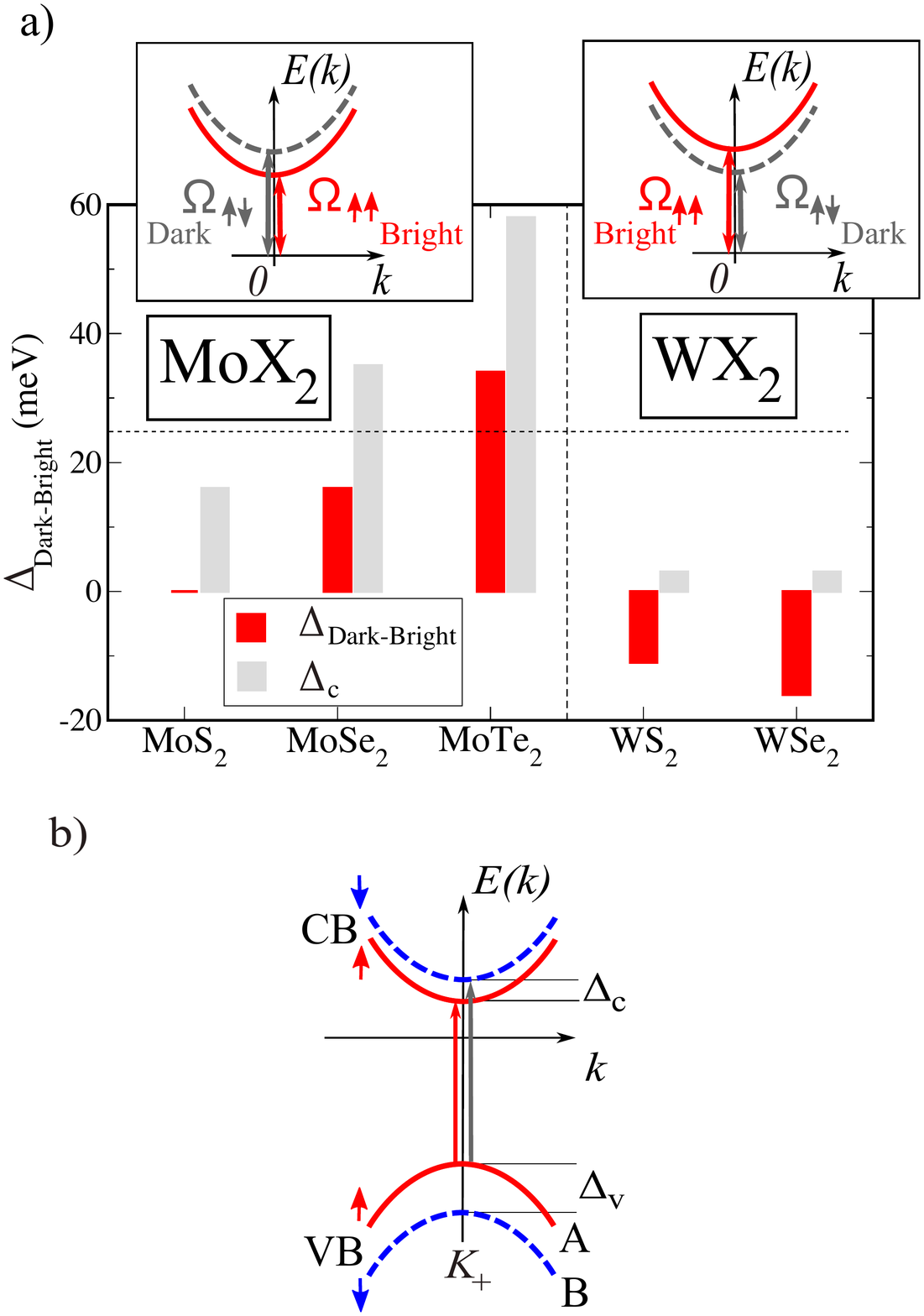}
\caption{\label{fig:fig1} a) Dark-bright energy splittings of the A $1s$ state excitons in the MoX$_2$ and WX$_2$ systems at $G_0W_0$+BSE level of theory,  the bright exciton energy being set to 0, in comparison with the conduction band splitting induced by spin-orbit coupling $\Delta_{\textrm c}$. The sign of the $e$-$h$ exchange contribution is negative. 
Insets are the corresponding qualitative band structure of the excitons in MoX$_2$ (left) and WX$_2$ (right), with bright and dark excitons in red and dotted gray respectively. b) Schematics of the one particle band structure of systems like MoX$_2$ ML in the $K_+$ valley, the $K_-$ valley can be obtained by time-reversal symmetry. The spin-orbit induced splittings for valence ($\Delta_{\textrm v}$) and conduction bands ($\Delta_{\textrm c}$) are shown.}
\end{figure} 
%%%%%%%%%%%%%%%%%%%%%%%%%%%%%%%%%%%%%
\indent Figure~\ref{fig:fig1}b) presents the schematics of a typical TMDC ML band structure in a single particle picture. In addition to the large spin-orbit splitting $\Delta_{\textrm v}$ between A and B \textit{valence bands} (VBs), the interplay between inversion asymmetry and spin-orbit interaction also yields a smaller spin-splitting ($\Delta_{\textrm c}$) between the lowest energy \textit{conduction bands} (CBs) \cite{Kosmider:2013bj,Liu:2013bu,Kormanyos:2015a,Dery:2015a}. 
 First let us consider A-excitons, with the transitions involving only the highest VB energy (A) and the two lowest CBs ($\uparrow$ or $\downarrow$), see Figure~\ref{fig:fig1}a). Here $\Omega_{\uparrow\downarrow}$ represents dark (spin forbidden) transitions, and $\Omega_{\uparrow\uparrow}$ shows bright transitions.  As a first approximation, one could consider that the energy splitting  $\Delta_{\textrm{Dark-Bright}}=\Omega_{\uparrow\downarrow}-\Omega_{\uparrow\uparrow}$ between bright and dark excitons is mainly due to the CB splitting $\Delta_{\textrm c}=E_{\textrm{CB} \downarrow}-E_{\textrm{CB}\uparrow}$ induced by SOC.
Depending on the material, $\Delta_{\textrm c}$  can either be negative for WX$_2$ or positive for MoX$_2$ systems (X=S, Se), as standard functional theory (DFT) and tight-binding calculations have shown~\cite{Zhu:2011km,Kosmider:2013bj,Cheiwchanchamnangij:2013kk,Kosmider:2013bx,Liu:2013bu,Roldan:2014jr}.
However, this single particle approach misses a key ingredient for the accurate determination of the dark-bright exciton splitting: the short-range part of the $e$-$h$ Coulomb exchange interaction for the exciton~\cite{Salmassi:1989wj, Blackwood:1994vl,Amand:1997wm}. So far in the context of TMDC MLs only the effects of the long-range exciton exchange interaction on the valley dynamics of bright exciton states in TMDC monolayers have been studied~\cite{Yu:2014fw,Zhu:2014hr,Glazov:2015fk,Yu:2014a}.\\
\indent Here we demonstrate that the $e$-$h$ exchange interaction within the exciton causes a splitting between the low energy optically bright and dark excitons in TMDC MLs. We find a giant exchange term of the order of 20 meV, more than 100 times larger than in GaAs quantum wells for instance~\cite{Salmassi:1989wj, Blackwood:1994vl,Amand:1997wm}. For all the investigated MoX$_2$ and WX$_2$ MLs, this local field effect due to the exchange interaction is added algebraically to the CB spin-orbit term $\Delta_{\textrm c}$, yielding a splitting $\Delta_{\textrm{Dark-Bright}}$ between bright and dark exciton that differs significantly from $\Delta_{\textrm c}$, see comparison in Figure~\ref{fig:fig1}a). We find that for MoX$_2$ systems, A-exciton dark states are higher in energy than bright ones, whereas this order is reversed for WX$_2$ MLs. Interestingly for WSe$_2$ ML we obtain $\Delta_{\textrm{Dark-Bright}}=-16$ meV, a value in good agreement with a first experimentally derived value~\cite{Zhang:2015er}. Importantly, we also report results on the exciton states involving the B valence band, which are fully consistent when compared to $\Delta_{\textrm{Dark-Bright}}$ for the A exciton, since the conduction band splitting contributes with the opposite sign. Moreover we show that the determination of the CB spin-orbit splitting $\Delta_{\textrm c}$ requires to perform DFT-based calculations within the $GW$ approach~\cite{Hedin:1965tu,Aryasetiawan:1998wz}. Significant changes compared to standard DFT level calculations~\cite{MolinaSanchez:2013hz,Qiu:2013jv, Kosmider:2013bj,Kosmider:2013bx} are obtained for all TMDC MLs.\\
\indent \textit{Computational Details.---} The atomic structures, the quasi-particle band structures and optical spectra are obtained from DFT calculations using the VASP 
package \cite{Kresse:1993a,kresse:prb:96}. Heyd-
Scuseria-Ernzerhof (HSE) hybrid functional~\cite{heyd:jcp:04_a,heyd:jcp:05,paier:jcp:06} is used as approximation of the exchange-correlation electronic term, 
as well as the Perdew-Burke-Ernzerhof (PBE) one \cite{Perdew:1996prl}.  It uses the plane-augmented wave scheme \cite{blochl:prb:94,kresse:prb:99} to treat core electrons. 
Fourteen electrons for Mo, W atoms and six for S, Se ones are explicitly included in the valence states. All atoms are allowed to relax with a force convergence criterion below
 $0.005$ eV/\AA. After primitive cell relaxation, the optimized lattice parameters, given in Table \ref{tab:tab1}  are in good agreement (1\%) with previous calculations~\cite{Ramasubramaniam:2012a} with all values slightly 
 larger than the bulk experimental ones. A grid of 
12$\times$12$\times$1 k-points has been used, in conjunction with a vacuum height of 17 \AA, to take benefit of error's cancellation in the band gap estimates~\cite{Huser:2013a}, 
and to provide absorption spectra in reasonable agreement with experiments as suggested in different works~\cite{Klots:2014a, MolinaSanchez:2013a}. 
A gaussian smearing with a width of 0.05 eV is used for partial occupancies, when a tight electronic minimization tolerance  
of $10^{-8}$ eV is set to determine with a good precision the corresponding derivative of the orbitals with respect to $k$ needed 
in quasi-particle band structure calculations. Spin-orbit coupling was also included non-self-consistently to determine eigenvalues and wave functions as input for the full-frequency-dependent $GW$ 
calculations~\cite{Shishkin:2006a} performed at the $G_0W_0$ level but also at $GW_0$ level with 3 iterations of the $G$ term, when necessary. 
The total number of states included in the $GW$ procedure is set to 600, after a careful check of the direct band gap convergence, to be smaller than 0.1 eV.  All optical excitonic transitions have been calculated by solving the Bethe-Salpeter Equation as follows~\cite{Hanke:1979to,Rohlfing:1998vb}:
\begin{equation}
\left( \varepsilon_c^{\textrm{QP}}-\varepsilon_v^{\textrm{QP}}\right) A_{vc} + \sum_{v'c'} \bra{vc}K^{eh}\ket{v'c'}A_{v'c'}=\Omega A_{vc},
\end{equation}
where $\Omega$ are the resulting $e$-$h$ excitation energies. $A_{vc}$ are the corresponding eigenvectors, when $\varepsilon^{QP}$ are the single-quasiparticle energies obtained at the $G_0W_0$ level, and $K^{eh}$ being the CB electron-VB hole interaction kernel.  
This term consists in a first attractive screened direct term and a repulsive exchange part. Practically we have included the six highest valence bands and the eight lowest conduction bands to obtain eigenvalues and oscillator strengths on all systems. Dark excitons are characterized by oscillator strengths around one thousand times smaller than bright ones~\cite{note1}.

%%%%%%%%%%%%%%%%%%%%%%%%%%%%%%%%%%%%%%%%%%%%%%%%%%%%%%%%%
\begin{table}[htp]
\begin{center}
\begin{tabular}{ccccccccccc}
\hline
\hline
%&& \multicolumn{5}{c}{$\Delta_{\textrm{B-A}}$ (meV)} && \multicolumn{3}{c}{$\displaystyle \Delta_{\Omega_{\uparrow\downarrow}-\Omega_{\uparrow\uparrow}}$ (meV)}\\
%\cline{3-7} \cline{9-11}
%&& $G_0W_0$-PBE && $G_0W_0$-HSE && Exp. & & A-excitons && B-excitons\\
%\hline
%MoS$_2$ && 0.16 && 0.18 & & 0.16\footnote{Reference~\cite{Mak:2012fp}} & & 0~(-5) && -24~(-20)\\
%MoSe$_2$ && 0.21 && 0.24 && 0.22\footnote{Reference~\cite{Wang:2015gw}} & & 16~(11) && -45~(-37) \\
%MoTe$_2$ && 0.30 && 0.33 && 0.26\footnote{Reference~\cite{Ruppert:2014kd}} & & 34 (25) && -30 (-9) \\
%WS$_2$ && - && 0.45 && 0.38\footnote{Reference~\cite{Zhu:2015ex}} && -11 && -6  \\
%WSe$_2$ & & - & & 0.49 && 0.43\footnote{Reference~\cite{Wang:2015kb}} &&  -16 && -37\\ 

Monolayer && \multicolumn{3}{c}{$\Delta_{\textrm{B-A}}$ (eV)} && \multicolumn{3}{c}{$\displaystyle \Delta_{\textrm{Dark-Bright}}$ (meV)}\\
\cline{3-5} \cline{7-9}
&& $G_0W_0$-HSE && Exp. & & A-excitons && B-excitons\\
\hline
MoS$_2$ && 0.18  (0.16) & & 0.16\footnote{Reference~\cite{Mak:2012fp}} & & 0~(-5) && -24~(-20)\\
MoSe$_2$ && 0.24 (0.21) && 0.22\footnote{Reference~\cite{Wang:2015gw}} & & 16~(11) && -45~(-37) \\
MoTe$_2$ && 0.33 (0.30) && 0.26\footnote{Reference~\cite{Ruppert:2014kd}} & & 34 (25) && -30 (-9) \\
WS$_2$ && 0.45 && 0.38\footnote{Reference~\cite{Zhu:2015ex}} && -11 && -6  \\
WSe$_2$ && 0.49 && 0.43\footnote{Reference~\cite{Wang:2015kb}} &&  -16 && -37\\ 

\hline
\hline
\end{tabular}
\end{center}
\caption{$\Delta_{\textrm{B-A}}$ : calculated energy splitting between $1s$  A and B bright excitons, $\Delta_{\textrm{Dark-Bright}}$ : calculated dark-bright energy separation for various TMDC MLs of the $1s$ state excitons at the $G_0W_0$+BSE level using HSE orbitals, the values in parentheses are extracted from $G_0W_0$-PBE calculations.}
\label{tab:tab1}
\end{table}
%%%%%%%%%%%%%%%%%%%%%%%%%%%%%%%%%%%%%%%%%%%%%%%%%%%%%%%%%

\textit{Results and Discussion.---} Figure~\ref{fig:fig1}a) and Table~\ref{tab:tab1} present the calculated values of the splittings $\Delta_{\textrm{Dark-Bright}}$ between non-optically active $1s$ exciton transitions $\Omega_{\uparrow\downarrow}$ and active ones $
\Omega_{\uparrow\uparrow}$. This splitting is positive for A exciton in the cases of MoSe$_2$ and MoTe$_2$ MLs. These results are compatible with the measured dependence of the photoluminescence (PL) intensity with respect to the temperature~\cite{C5NR06782K,Wang:2015ko,Zhang:2015er,Withers:2015cf}. If we assume, that the direct contribution terms of the $K^{eh}$ kernel are very close for the two distinct spins orientations for one particular ML, the effect of the $e$-$h$ exchange term is to partially compensate the conduction band splitting, by substracting roughly 20 meV 
for those two systems from $\Delta_{\textrm c}$ to obtain the exciton dark-bright splittings, see Figure~\ref{fig:fig1}a). In other words, the CB spin-orbit splitting and the exciton exchange term have opposite signs. Interestingly the exciton exchange contribution seems to be an invariant quantity for the entire family, at least for the tested TMDC MLs, and agrees well with the single available value of this term in the literature, coming from calculations of a freestanding MoS$_2$ ML~\cite{Qiu:2015fa}. 
For the particular MoSe$_2$ and MoTe$_2$ cases, since the conduction band splitting is larger than the exciton exchange term, $\Delta_{\textrm{Dark-Bright}}$ is positive, as shown in Figure~\ref{fig:fig1}a). Local-field effects (exciton exchange term) also act on the B exciton, accordingly to the sign change of $\Delta_\textrm c$ for the transition involving the B valence band series but with almost the same repulsive exchange term, now the $\Omega^B_{\uparrow\downarrow}$ transition appears below the $\Omega^B_{\uparrow\uparrow}$ one. Note that for those two systems, there is also a change in the direct term, since it appears less attractive than in the A case, this is particularly true for MoTe$_2$.\\ 
\indent The MoS$_2$ ML is very intriguing, our calculations show that the CB spin-orbit splitting and the exciton exchange term almost cancel each other yielding a very small bright-dark exciton separations. As a 
consequence, $\Delta_{\textrm{Dark-Bright}}$ is negative for PBE-based calculations (-5 meV) and 0 meV when HSE orbitals are used. This is in line with the previous determination of Qiu {\textit{et al}} \cite{Qiu:2015fa} for the same system, since $\Omega^A_{\uparrow\downarrow}$ is lower in energy than $\Omega^A_{\uparrow\uparrow}$ by approximatively 17 meV and $\Omega^B_{\uparrow\downarrow}$ by around 23 meV with respect to $
\Omega^B_{\uparrow\uparrow}$.  We observe the same trend, the difference being certainly due to our larger $\Delta_\textrm c$ values obtained at the $GW$ level, see below. On the B exciton, we also observe a small 
decrease of the direct contribution that makes the absolute energy difference between the parallel and anti-parallel spins slightly smaller that the simple addition of the 
conduction band splitting and the exciton exchange term.

%%%%%%%%%%%%%%%%%%%%%%%%%%%%%%%%%%%%%
\begin{figure}[htp]
\includegraphics[width=0.49\textwidth]{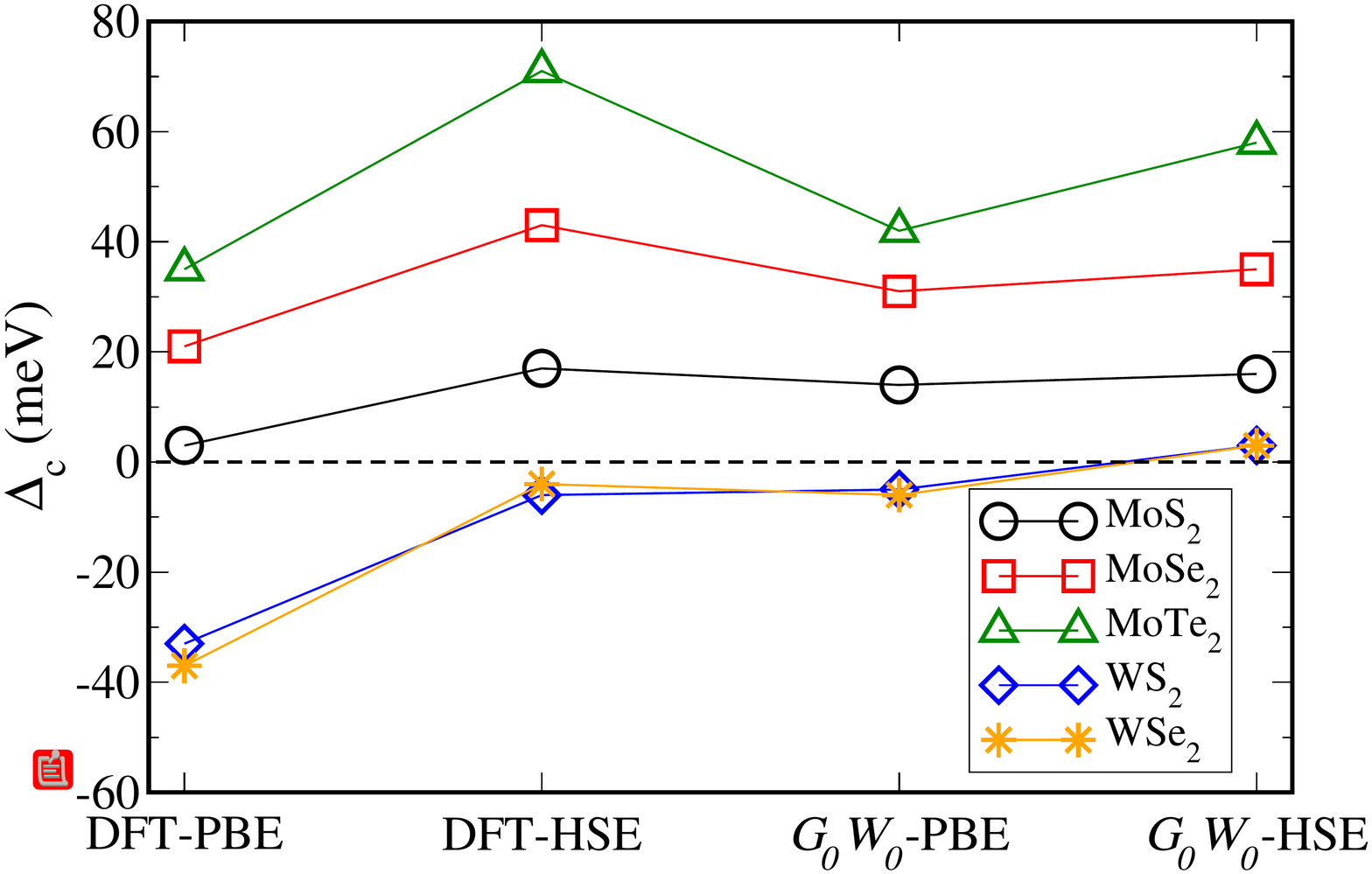}
\caption{\label{fig:fig2} Evolution of the calculated conduction band splittings with respect to the level of theory for five TMDC monolayers. The data are extracted from Table SI-1.}
\end{figure} 
%%%%%%%%%%%%%%%%%%%%%%%%%%%%%%%%%%%%%

\indent For WS$_2$ and WSe$_2$ MLs, since $\Delta_\textrm c$ is very small in both cases, see Figure~\ref{fig:fig1}a) and Supplemental Material~\cite{supp-mat}, the exchange term contribution dictates the bright-dark exciton splitting inducing the bright exciton to lie at lower energy compared to the dark one. The corresponding energy separations remain very modest when compared to the binding energy, but at the same time allow to actively populate the bright excitons when increasing the sample temperature. To be more specific for WSe$_2$, we propose here an estimate of the dark-bright energy splitting smaller than 20 meV, in good agreement with the indirect determination based on a recent fit procedure of experimental data~\cite{Zhang:2015er}.\\ 
\indent The determination of the bright and dark exciton splitting in TMDC MLs, requires very high accuracy in the calculations of spin-orbit splitting in the \textit{conduction bands}. 
{It is well documented that standard DFT fails predicting band structure parameters like band gap or effective masses. Due to the ground state character of the DFT, conduction band splittings need to be more accurately described by means of many-body based calculations. Here our $GW$-based calculations provide more accurate band structure parameters compared to standard DFT, as it has been shown for similar systems~\cite{Ramasubramaniam:2012a,Zhuang:2013gk,MolinaSanchez:2013a,Klots:2014a} Here we compare 
the SOC effects on both valence and conductions bands, by means of standard DFT with our results from more advanced $GW$ schemes. All data corresponding to VB energy splittings $\Delta_\textrm v$  values are reported in Supplemental Material~\cite{supp-mat}. Those values agree very well with previous DFT-based reports~\cite{Ramasubramaniam:2012a,Kosmider:2013bj,Cheiwchanchamnangij:2013kk,Kosmider:2013bx}. It clearly shows that adding a partial exact exchange contribution to the electronic exchange-correlation functional significantly enlarges $\Delta_\textrm v$ for all TMDC MLs. This trend is directly observable in the computed absorption spectra and more specifically in the $\Delta_{\textrm{B-A}}$ results of Table~\ref{tab:tab1} which include contributions both from $\Delta_\textrm v$ and $\Delta_\textrm c$ but also $e$-$h$ interaction effects. HSE-based calculations always give larger A-B splittings than the experimental estimates, but still remain in a reasonable agreement with them. 
Considering $\Delta_\textrm c$, at the DFT level, our results are similar to previous reports~\cite{Kosmider:2013bj,Kosmider:2013bx}: for the MoX$_2$ family, when going from a semi-local to a hybrid approximation of the electronic exchange-correlation term, the splittings between spin-$\uparrow$ and $\downarrow$ CB states are enhanced, see Figure~\ref{fig:fig2} and Supplemental Material~\cite{supp-mat}. Interestingly for the WS$_2$ and WSe$_2$ cases, if now $\downarrow$-CB is the lowest unoccupied state, the energy separation is significantly reduced in hybrid functional calculations, meaning that short-range Hartree-Fock term strongly influence both atomic SOC contributions from the transition metal and from the chalcogen, to make them competitive with opposite signs~\cite{Kosmider:2013bj}. 
For the MoS$_2$ and MoSe$_2$ cases, our $G_0W_0$ calculations based on PBE or HSE wave functions provide similar $\Delta_\textrm c$ splittings suggesting a non-dependence band splitting on the ground state orbitals in contrast with MoTe$_2$ ML, probably due to an overestimation of the chalcogen contribution using HSE-based orbitals, which leaves $\Delta_\textrm c$ 40\% larger even after applying a single shot $GW$ correction. 
Remarkably for W-based systems the $GW$ scheme reduces drastically $\Delta_\textrm c$ starting from PBE orbitals, and even reverses again the spin-up and spin-down conduction bands ordering in HSE-based calculations. This clearly indicates that $\uparrow$ and $\downarrow$ CB states are almost degenerate at the $GW$ level, contrary to standard DFT predictions.\\
\indent The $G_0W_0$ results reveal further interesting details: As it has been already reported~\cite{Ramasubramaniam:2012a} and mentioned in the present letter, all the extracted values of $GW$ calculations, namely $\Delta_\textrm v$, $\Delta_\textrm c$ and $E_\textrm g$ (free carrier gap value), are very sensitive to the choice of starting set of orbitals. 
Indeed as shown in the Supplemental Material~\cite{supp-mat}, larger values (+0.1eV) of $E_\textrm g$ are yielded when HSE wave functions are used. 
For WS$_2$ and WSe$_2$, PBE calculations give an indirect band gap in $K_+$-$\Lambda$, with values smaller by 0.03 and 0.05 eV respectively when compared to direct $K_+$-$K_+$ gaps, in agreement with a recent experimental result~\cite{Zhang:2015js}.
Note that these discrepancies remain even after two additional iterations in a $GW_0$ scheme, pointing out the importance of the starting wave functions for W-based systems. The choice of lattice parameter value is crucial for the resulting band structure properties: using the experimental lattice parameter results in drastically different values for $\Delta_\textrm c$, $\Delta_\textrm v$ and $E_\textrm g$~\cite{supp-mat}.\\
\indent \textit{Conclusions.---} Our calculations demonstrate that the bright to dark exciton energy splitting in MoX$_2$ and WX$_2$ monolayers depends in both sign and amplitude on the strong local-field effects in the $e$-$h$ interaction, or in other words, the electron-hole short range Coulomb exchange within the exciton. In addition to this exchange energy, the conduction band spin-orbit splitting $\Delta_c$ needs to be taken into account, which has initially been put forward as the main origin of the dark-bright exciton splitting. To measure the splitting between $\Delta_c$ between electron spin states in the absence of holes, techniques like angle resolved photoemission spectroscopy (ARPES) \cite{Zhang:2014a} or electron spin resonance would be desirable, as pure optical spectroscopy techniques (absorption, emission) cannot separate this contribution from the strong electron-hole Coulomb exchange effects.\\

We thank Misha Glazov for fruitful discussions and ANR MoS2ValleyControl and Programme Investissements d'Avenir ANR-11-IDEX-0002-02, reference ANR-10-LABX-0037-NEXT for financial support. The authors also acknowledge the CALMIP initiative for the generous allocation of computational times, through the project p0812, as well as the GENCI-CINES, GENCI-IDRIS and GENCI-CCRT for the grant x2014096649. I. C. Gerber also thanks the CNRS for his financial support. X. M. acknowledges the Institut Universitaire de France. B.U. is supported by ERC Grant No. 306719.

\bibliography{Dark-bright-bib}% Produces the bibliography via BibTeX.

\begin{thebibliography}{71}
\bibitem[{not({\natexlab{a}})}]{note0}
\bibinfo{note}{Here we work in the electron formalism (not hole), where the
  conduction electron has the same spin as the valence state it occupied
  before. Same spin transitions are therefore optically allowed.}
\end{thebibliography}
\end{document}